\documentclass[10pt,conference]{IEEEtran}
\usepackage[utf8]{inputenc}

\usepackage{soul}
\usepackage{xcolor,colortbl}
\usepackage{xspace}
\PassOptionsToPackage{hyphens}{url}
\usepackage{hyperref}
\usepackage{graphicx}
\usepackage{listings,multicol}
\usepackage[caption=false]{subfig}  
\usepackage[export]{adjustbox}
\usepackage{amsmath}
\usepackage{textcomp}
\usepackage{listings}
\usepackage{cleveref}
\usepackage{algorithm}
\usepackage[noend]{algpseudocode}
\usepackage{float}
\usepackage{svg}
\usepackage[justification=centering]{caption}
\usepackage{mdframed}
\usepackage[switch]{lineno}

\usepackage{booktabs}
\usepackage{multirow}
\usepackage{graphicx}

\definecolor{maxc}{RGB}{201,250,190}
\definecolor{minc}{RGB}{255,206,173}

\newcommand{\woc}{WoC\xspace}

\graphicspath{ {./figures/} }

\begin{document}
\title{Constructing Temporal Networks of OSS Programming Language Ecosystems}

\author{\IEEEauthorblockN{Alexander Agroskin}
\IEEEauthorblockA{\textit{Weizmann Institute of Science} \\
Israel \\
sasha.agroskin@gmail.com}
\and
\IEEEauthorblockN{Elena Lyulina}
\IEEEauthorblockA{\textit{JetBrains Research} \\
Serbia\\
elena.lyulina@jetbrains.com}
\and
\IEEEauthorblockN{Sergey Titov}
\IEEEauthorblockA{\textit{JetBrains Research} \\
Cyprus \\
sergey.titov@jetbrains.com}
\and
\IEEEauthorblockN{Vladimir Kovalenko}
\IEEEauthorblockA{\textit{JetBrains Research} \\
The Netherlands \\
vladimir.kovalenko@jetbrains.com}
}

\maketitle
\thispagestyle{empty}
\pagestyle{empty}
\begin{abstract}

One of the primary factors that encourage developers to contribute to open source software (OSS) projects is the collaborative nature of OSS development.
However, the collaborative structure of these communities largely remains unclear, partly due to the enormous scale of data to be gathered, processed, and analyzed.  

In this work, we utilize the World Of Code dataset, which contains commit activity data for millions of OSS projects, to build collaboration networks for ten popular programming language ecosystems, containing in total over 290M commits across over 18M projects.
We build a collaboration graph representation for each language ecosystem, having authors and projects as nodes, which enables various forms of social network analysis on the scale of language ecosystems.
Moreover, we capture the information on the ecosystems' evolution by slicing each network into 30 historical snapshots.
Additionally, we calculate multiple collaboration metrics that characterize the ecosystems’ states.

We make the resulting dataset publicly available\footnote{\label{footnote:reppackage}Reproducibility package: \url{https://doi.org/10.5281/zenodo.6414660}}, including the constructed graphs and the pipeline enabling the analysis of more ecosystems.

\end{abstract}

\newcommand{\etal}{\emph{et al.}\xspace}
\section{Introduction}\label{sec:introduction}
Over the last few years, research on open source software (OSS) development has become more active~\cite{cosentino2017systematic}.
Companies like DigitalOcean and StackOverflow publish yearly reports about the state of OSS development~\cite{stackoverflow}, and several academic studies review the state of OSS hosting platforms like GitHub~\cite{coelho2020github, cosentino2017systematic}.
Research of OSS development is crucial not only for understanding how the OSS community functions, but also for making informed decisions about development plans for big projects~\cite{sbai2018exploring} and for policy-making~\cite{lerner2010comingled}.

Open source software development is often collaborative.

Collaboration of OSS developers has been extensively studied on the small scale.
Numerous studies have focused on collaboration in individual projects, such as the Linux kernel~\cite{tan2020scaling},
or small samples of projects from GitHub~\cite{ortu2018mining}. 
Such works shed light on the collaboration process in individual small teams and projects and help the project owners to organise the development in a more efficient way~\cite{tan2020scaling}. 

Several existing studies are focused on collaboration networks of big codebases~\cite{kabakus2020githubnet},
or sets of projects united by one technology~\cite{dey2018software}.
The bigger scale of such studies helps the researchers to test the applicability of general social science phenomena to the OSS community. 
However, to our knowledge, no existing studies explored collaboration on the scale of entire programming language ecosystems. 
One possible reason for the lack of such studies is the technical challenge of processing such a large volume of collaboration data. 

Research on the scale of the language ecosystems is promising, as it presents an opportunity to understand collaboration on a larger scale. 
Programming language ecosystems are a rare example of enormous collaborative communities united by common context and knowledge.
Similarly to how studies of collaboration in scientometrics yielded important ideas such as theories of citation~\cite{mingers2015review},
large-scale studies of collaboration in OSS could benefit the software engineering research community.
The choice of programming language is listed among the factors that best explain the growth of OSS projects in terms of new contributions~\cite{fronchetti2019attracts}.
Therefore, considering ecosystems of programming languages independently enables reasoning about the particulars of collaboration structure and patterns inherent to these languages, which might be crucial for both developers and those responsible for language development.

In this work, we focus on developer collaboration in \emph{programming language ecosystems} --- socio-technical networks of projects and authors united by using the same programming language.
To expand the scope of existing research of collaboration in software engineering,
we construct a temporal collaboration network and trace the evolution of the ecosystem for each of the ten selected popular programming languages.
To build such networks and overcome the challenges of collecting data, we use the World of Code dataset~\cite{ma2019world}, which contains full history of contributions to almost all OSS projects hosted on multiple platforms.
The total computation time to process this data and build collaboration networks exceeded 170 hours using a standard 4-core Intel processor with 32 GB of RAM.
The resulting network dataset with full collaboration history for ten programming language ecosystems along with multiple collaboration metrics is publicly available on Zenodo\textsuperscript{\ref{footnote:reppackage}}. Additionally, we provide an interactive demonstration to inspect and compare the metrics' values\footnote{\label{footnote:demo}Interactive demonstration: \url{https://datalore.jetbrains.com/view/report/Gr9eBMPW1BGe58l2Y3SqrB}}. 

We believe that our work will be useful for other researchers analysing the collaborations within OSS development, as well as for the core language developers. The provided dataset and metrics could help to determine the current ecosystem state, while historical slices might be used to analyze the ecosystem changes and predict further evolution. 
Contributor-centered organization of the data enables comprehensive analysis of social networks of collaborators. 
By performing all the technically challenged and time-consuming operations, we expanded the field of research it might be applicable to, including social and collaboration science. 

\section{Background and related work}\label{sec:background}

\subsection{Studies on Collaboration In OSS}

Several notable works formulated the approach to construction of collaboration networks and use of network properties~\cite{lopez2004applying, xu2005topological}. 
Lopez~\etal suggested the method to build a collaboration network from version control systems data and proposed a set of network properties for analysis. 
They found that collaboration networks of these projects fit to the small-world network model. 
The same findings were revealed by Xu~\etal\cite{xu2005topological} on a bigger scale --- in their analysis, collaboration networks of more than 80K projects from SourceForge also demonstrated the properties of small-world networks. 
They argue that dense structure of collaboration networks ensures efficiency of the open source development.

The more recent studies of collaboration networks look deeper into the collaboration data, searching for various social and organisational phenomena~\cite{tan2020scaling, gote2019git2net}. 
A notable work involving collaborative network analysis on a project scale is done by Gote \etal~\cite{gote2019git2net} with git2net --- a toolkit for constructing temporal co-editing networks from Git history. 
Such insights are useful for the organisation of open source projects: knowing how collaboration structure is linked to the work process and specifics of OSS contribution dynamics could help set up an efficient development process.
\subsection{Language ecosystem datasets}
There exist several works that leverage datasets with almost all publicly available source code for particular programming languages ~\cite{ray2014large, bahrami2021pytorrent}. 
However, such datasets are not designed for reasoning about the collaborations within language ecosystems, which requires not only code but also information about authors and version control history. 
Several datasets contain most of this data for multiple programming languages: GH Archive~\cite{gharchive}, Software Heritage dataset~\cite{di2017software}, GHtorrent~\cite{gousios2012ghtorrent} and \woc~\cite{ma2019world} being the most complete data sources of the whole OSS.
The latter, collected and maintained by Ma \etal, features open-source projects available on major VCS hosting platforms, combined into a dataset with 12B git objects including information about projects, authors, files, and their historical changes.
This information should be sufficient to build network representations of various OSS ecosystems.
For example, Mockus \etal~\cite{mockus2020deforking} build a graph representation of the \woc data to run community detection algorithms. 
As a result, they found that the biggest cluster of projects contains around 400K interconnected repositories.

However, to the best of our knowledge, there are no works aimed at building collaboration networks for multiple ecosystems of programming languages at once, especially together with their historical changes over time.
While being a technically challenging task, building these networks enables reasoning about collaborative structure and patterns inherent to each language ecosystem, highlighting the unique features of each language and its community.

\section{Methodology}\label{sec:methodology}
\subsection{Ecosystem selection}\label{subsec:choosingeco}
Our work focuses on OSS ecosystems that encompass programming language communities.
We use the definition of an ecosystem by Manikas \etal~\cite{manikas2016revisiting} --- ecosystems are \emph{holistic networks of projects and contributors united by the use of the same technology}.
In these ecosystems, we consider every type of project: software, educational, data storage, and others, as they all contribute to the ecosystem's structure.

The main source of data for our work is the World of Code project~\cite{ma2019world}.
It provides information on every collected commit, including the commit's author, project, timestamp, and programming languages of the files affected by the commit. 
For this paper, we use the version of the dataset that includes all data identified by the WoC project until February 12, 2021.

The first step of our work is choosing the target language ecosystems to be included in the data.
However, not all of the ecosystems in \woc are sufficiently large or active to be of interest when constructing a network dataset. 
Because of this, we select language ecosystems with a total number of commits in WoC exceeding \textbf{one million}.
Because of limited computational power, we could not analyse some of the most popular programming languages such as Python and JavaScript. 
Additionally, due to an overlap in file extensions of C and C++ source code, distinguishing these two ecosystems would require an analysis of the entirety of the source files, which is beyond our computational resources.
 
Ultimately, we select 10 programming language ecosystems to be included in the dataset.
In \Cref{tab:basic_counts}, we present statistics for each of the chosen languages ecosystems.

\begin{table}[h!]
\centering
\begin{tabular}{@{}lllll@{}}
\toprule
 {}          & \textbf{Projects} & \textbf{Authors} & \textbf{\begin{tabular}[c]{@{}l@{}}Authors\\ (dealiased)\end{tabular}} & \textbf{Commits} \\ \toprule
\multicolumn{1}{l}{\textbf{Java}}   & 9.61M             & 7.59M            & 6.02M                                                                  & 157.7M           \\ \midrule
\multicolumn{1}{l}{\textbf{C\#}}    & 3.78M             & 2.94M            & 2.35M                                                                  & 53.9M            \\ \midrule
\multicolumn{1}{l}{\textbf{Ruby}}   & 2.36M             & 1.48M            & 1.14M                                                                  & 25.8M            \\ \midrule
\multicolumn{1}{l}{\textbf{Go}}     & 0.85M             & 0.77M            & 0.6M                                                                   & 22.9M            \\ \midrule
\multicolumn{1}{l}{\textbf{Kotlin}} & 0.61M             & 0.45M            & 0.38M                                                                  & 7.9M             \\ \midrule
\multicolumn{1}{l}{\textbf{Perl}}   & 0.36M             & 0.36M            & 0.31M                                                                  & 7.6M             \\ \midrule
\multicolumn{1}{l}{\textbf{Rust}}   & 0.33M             & 0.31M            & 0.26M                                                                  & 7.4M             \\ \midrule
\multicolumn{1}{l}{\textbf{Scala}}  & 0.22M             & 0.23M            & 0.18M                                                                  & 7M               \\ \midrule
\multicolumn{1}{l}{\textbf{R}}      & 0.63M             & 0.5M             & 0.42M                                                                  & 5.6M             \\ \midrule
\multicolumn{1}{l}{\textbf{Dart}}   & 0.37M             & 0.25M            & 0.21M                                                                  & 3.6M             \\ \bottomrule
\end{tabular}
\smallskip
\caption{Basic statistics for the language ecosystems featured in this study.}
\label{tab:basic_counts}
\end{table}

\subsection{Slicing temporal data}\label{subsec:slicing}

After selecting the ecosystems, we proceed with extracting temporal information by slicing the data on each ecosystem into subsets, each corresponding to a time period. 
Because we are interested in capturing the dynamics of ecosystem networks for further comparison, we need to have the same number of slices for each ecosystem.
At the same time, we need to avoid creating slices that are too small, during which the chosen ecosystem is static.

To this end, we use the following slicing metholodology: each slice is made when $\frac{1}{N}$ of the total commit count has occurred. 
We use $N=30$ as a middle ground between the computationally expensive higher rate of discretization and the less informative lower rate.
Thus, a slice is made when $3.33\%$ of the total commit count has occurred.
However, because of the high activity density in mature ecosystems, the distribution of commits is highly non-uniform and grows exponentially. 
Thus, we have set a minimum time period $T$ which a slice must span: in our case, following a common practice in social link studies \cite{DUNBAR201539}, it is a minimum time span of 6 months.
This helps us look at ecosystems with more granularity when the rate of change is high.

This approach solves both of our problems: (1) we guarantee that for every ecosystem there is an equal number of slices, and (2) each slice is guaranteed to be non-static.
While the chosen slice points correspond to points in time, caution should be exercised when viewing the slices as even evolution units, due to the length of the time periods being affected by the commit frequency.
However, this notion of ``commit-based time'' has been actively used before, such as by Lewis \etal~\cite{lewis2013does}.

\subsection{Graph construction}\label{subsec:graphconst}

For the purposes of network analysis, we construct an undirected graph for each slice of the ecosystem. We create nodes of two types: \emph{author nodes} and \emph{project nodes}. 
These nodes denote the authors and the projects attached to a commit in a slice of an ecosystem, and both are uniquely identified by the MD5 hash of their name in the WoC data.

To merge different version control system (VCS)  aliases of the same contributors we use the author deduplication maps provided by WoC, the construction of which is explained in detail in the work of Fry \etal~\cite{dedup}.
Additionally, because of a standard open-source contribution procedure of forking the project and opening the pull request using the commits to the fork, we use the ``fork-normalized'' project identifiers provided by the WoC to take into account this form of collaboration.
Deforking is done by applying community detection algorithms based on shared commits \cite{mockus2020deforking}, with all references to the forked projects replaced by references to the project at the root of the fork tree.

After constructing the project and author nodes, we construct two types of relationships:

A \emph{contribution edge} connects an author $A$ and a project $P$ if and only if there is a commit $C$ in a slice that has been authored by $A$ and created on project $P$ (or one of its forks).

A \emph{collaboration edge} connects two authors $A_1$ and $A_2$ if and only if there exist commits $C_1$ and $C_2$ (authored by $A_1$ and $A_2$ respectively) to a project $P$, and if $C_1$ and $C_2$ include changes to the same files.
As there are files with an abnormally large number of contributors, such as coding tasks where thousands of people work on the same files, we discard files that have more authors than $99.99\%$ of all files in a given slice, as these are unlikely to be a meaningful indicator of collaboration. 
We assume that the resulting links indicate that two authors may have communicated, collaborated, and shared knowledge.

\subsection{Choosing the metrics}\label{subsec:choosingmetrics}

For each constructed graph, we calculate several metrics. We aim to not only collect basic quantitative metrics but also use the graph nature of our data to explore the network metrics of the graph. To this end, we have selected a number of metrics, for both the entire graph and individual nodes.

\subsubsection{Metrics For The Entire Graph}\hfill

\textbf{Author} and \textbf{Project counts:} represent the number of authors and projects in a graph, respectively. 
They reflect the size of an ecosystem slice and provide a baseline to normalize other metrics for comparisons between languages with vastly different levels of popularity.

\textbf{Collaboration} and \textbf{Contribution Counts:} represent the total number of edges of the two types. 
These metrics can as well be used for a general estimation of ecosystem size.

\textbf{Component sizes:} a list of node counts of each connected component of a graph. 
The distribution of these sizes and especially the \textbf{size of the largest component} relative to the size of the entire graph, can serve as a measure of overall network connectivity.

\textbf{Collaboration} and \textbf{Contribution densities:} denote the fractions of all possible collaborations and contributions respectively that are present in our set of edges. 
These metrics show how close the graph is to a complete one, and can likewise be viewed as a measure of graph connectivity.

\subsubsection{Metrics for individual nodes}\hfill

\textbf{Collaboration} and \textbf{Contribution degrees:} the number of edges of a certain type connected to the node. 
As collaboration edges only connect authors with authors, collaboration degree is skipped for the project nodes.
These metrics can also be used as a measure of node importance --- \emph{degree centrality}~\cite{wasserman1994social}.

\textbf{Betweenness centrality:} in its precise form, this node importance metric is a sum of fractions of shortest paths between each pair of vertices that contain a given node.
Because finding the shortest paths between each pair of nodes is extremely computationally intensive on our scale, we use an implementation of Brandes' betweenness centrality approximation algorithm \cite{brandes2007centrality} provided by the Neo4j Graph Data Science library\footnote{Neo4j GDS Betweenness centrality: \url{https://neo4j.com/docs/graph-data-science/current/algorithms/betweenness-centrality/}}.

\textbf{Local clustering coefficients:} a metric quantifying how close is a given node to forming a complete graph with its neighbours.
We calculate this metric only for the author nodes and collaboration edges, as there are no relations between projects in our model, and thus a complete graph is impossible.
Clustering coefficient $C_n$ of a node $n$ with a degree $d_n$ and a number of triangles (complete graphs on three vertices) $T_n$ containing the node $n$ can be calculated as

$$C_n = \frac{2\,T_n}{d_n\,(d_n-1)}$$

Our choice of metrics is based on common network analysis practices.
Betweenness centrality and the clustering coefficient are used by Lopez \etal~\cite{lopez2004applying} specifically for OSS developer network analysis. 
The clustering coefficient can be used for community detection \cite{clusteringgenetic} and as a measure of risk in financial networks \cite{clusteringbanking}.
Degree and betweenness centrality are widely used in social network analysis \cite{maharani2014degree} and scientometrics \cite{leydesdorff2007betweenness}.
Connected components, especially the largest connected component, are important for measuring the network's ability to propagate information among developers, similarly to how these metrics are used in  epidemiology \cite{danon2011networks}. 

\subsection{Technical details}\label{subsec:processing}

The graphs are exported as database archives and are provided as part of the reproducibility package\textsuperscript{\ref{footnote:reppackage}}.
The resulting dataset consists of 10 parts, each corresponding to a single language. 
Each part contains 30 extracted slices as Neo4j database dumps, a TSV file with per-node metrics for every slice, and a JSON file with per-network metrics for every slice.
Additionally, we provide a JSON file with component size distribution for all slices of each ecosystem.

To use the dataset, import the Neo4j dump using the built-in tool. 
This results in a graph database with two types of nodes: $\texttt{PROJECT}$ and $\texttt{AUTHOR}$, and two types of edges: author-project edges $\texttt{CONTRIBUTED\_TO}$ and author-author edges $\texttt{COLLABORATED}$.
More detailed notes on usage, along with all the scripts used in the dataset creation are available in the reproducibility package\textsuperscript{\ref{footnote:reppackage}}.

\section{Dataset demonstration and Future work}\label{sec:futurework}
To showcase our dataset, we have built an interactive demonsration page\textsuperscript{\ref{footnote:demo}}, allowing the viewers to explore collaboration metrics for various slices and languages. 
We hope that this work spurs further research into large scale OSS analysis and social networks with utilization of our dataset:

\textbf{Programming language analysis}
This dataset may be used to investigate the consequences of language evolution on the project development, be it the release frequency, new language features, or an overall change in language popularity. 
Another avenue lies in the deeper analysis of the languages that fall into similar categories or have related use cases.

\textbf{Cluster analysis} 
Among other metrics, we include largest component size and clustering coefficients for each programming language.
One can go further and apply clustering techniques to the temporal graph networks.
Such study will help to detect stable clusters and reason about the process of ecosystem emergence and communities within it.

\textbf{Link prediction}
Another way of utilizing a set of networks' time slices is to predict the future appearance or disappearance of links between nodes, be it a collaboration between two authors or a contribution to a project. 

\textbf{Key nodes}
The analysis of the key nodes (e.g. nodes with high centrality) in each ecosystem can shed light on the ways the most influential authors and projects for each language affect the open source community.

\textbf{Comparison with the developer survey results}
A common approach for evaluating the overall state of software development is conducting developer surveys, gathering comprehensive results with a rather small but diverse sample of participants~\cite{jbecosystemstate}.

\begin{figure}[h]
    \centering
    \includegraphics[clip,width=\linewidth]{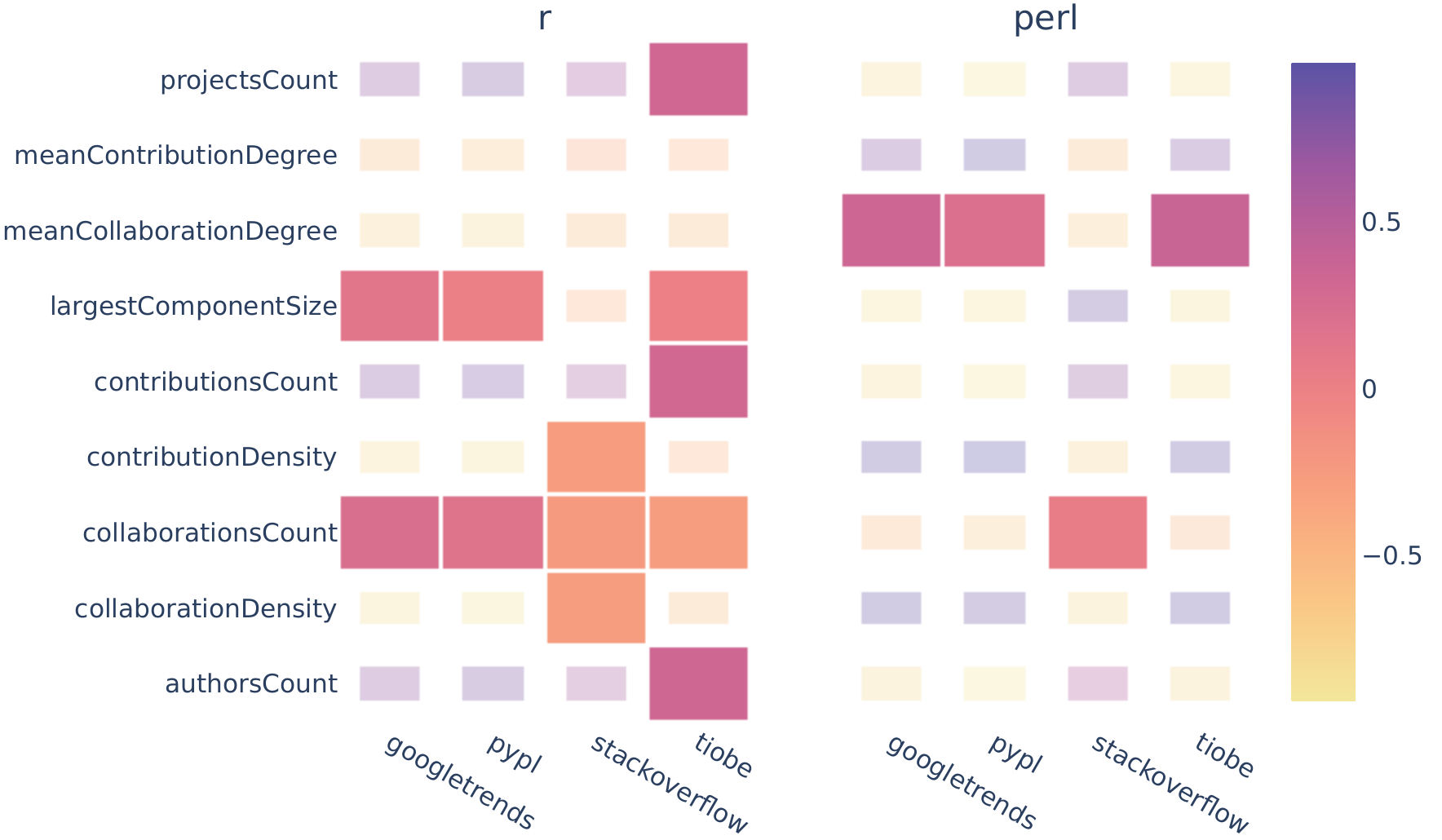}
    \caption{Correlation matrix of network metrics and language popularity metrics. Bigger cells indicate $p$-values $< 0.05$}
    \label{fig:correlation}
\end{figure}

Our dataset may serve both as a method of survey validation as well as an additional source of information.
To illustrate it, we explore languages' popularity and its correlation with collaborative metrics (\Cref{fig:correlation}). 
More details can be found on the demonstration page\textsuperscript{\ref{footnote:demo}}. 

\section{Threats to Validity}\label{sec:threats}

There are possible detrimental factors stemming from the original \woc dataset and the nature of open-source software in general.
\woc collects data from multiple sources and may include author and project duplication.
We are overcoming this challenge by using \woc deduplication maps, although, being heuristic-based, they might be error-prone.
Commit timestamps can likewise be a source of errors, as they depend on the possibly inaccurate local machine clock.
We alleviate this problem by disregarding commits that have a timestamp beyond the dataset collection date, but precise detection of these commits is beyond the scope of this paper.
Since public git repositories are not only used for collaborative software development, but also for online course platforms and knowledge storages, this may introduce noise to analysis of OSS development using \woc data. 

Our file-based approach to detecting collaborations does not reflect all the real-life steps such as forking a repository and opening pull requests due to lack of various metadata and inability to distinguish forks made for collaboration.
Another important consideration is our slicing approach, since the choice of these time periods is based not on a fixed time scale, but on the rate of commits made to an ecosystem.
Among other advantages, these slices still reflect the dynamics of the ecosystems, but caution should be exercised when transitioning from our commit-based slices to the standard notion of time.

\section{Conclusion}\label{sec:conclusion}
In this paper, we present and detail an approach to constructing a temporal network dataset from OSS commit information.
Our main contribution is a dataset consisting of 10 sliced networks, each describing the evolution of collaboration and contribution in an OSS ecosystem of a programming language.
The dataset is public and can be obtained from our reproducibility package on Zenodo\textsuperscript{\ref{footnote:reppackage}}. Additionally, we calculate a number of quantitative and network-based metrics for each slice of the dataset, allowing for a temporal view of some ecosystem characteristics on a large scale. To showcase possible applications of these metrics, we provide an interactive demo where one can inspect and compare various metrics for all provided languages. 
\section{Acknowledgements}\label{sec:ack}
We thank Audris Mockus and Alexander Nozik for their help and constructive feedback.

\bibliographystyle{IEEEtran}
\bibliography{IEEEabrv,paper}

\end{document}